\documentclass[5p,twocolumn,times,number]{elsarticle}
\usepackage{amsmath}
\usepackage{amssymb}
\usepackage{graphics}
\usepackage{graphicx}
\usepackage{epsfig}
\usepackage{dcolumn}
\usepackage{bm}

\begin{document}
\begin{frontmatter}

\title{Characterizations of GEM detector prototype}
\author[label3]{Rajendra~Nath~Patra}
\author[label1]{Amit~Nanda}
\author[label4]{Sharmili~Rudra}
\author[label1]{P.~Bhattacharya}
\author[label1]{Sumanya~Sekhar~Sahoo}
\author[label1]{S.~Biswas\corref{cor}}
\ead{saikat.ino@gmail.com, s.biswas@niser.ac.in, saikat.biswas@cern.ch}
\author[label1]{B.~Mohanty}
\author[label3]{T.~K.~Nayak}
\author[label2]{P.~K.~Sahu}
\author[label2]{S.~Sahu}

\cortext[cor]{Corresponding author}

\address[label1]{School of Physical Sciences, National Institute of Science Education and Research, Jatni - 752050, India}
\address[label2]{Institute of Physics, Sachivalaya Marg, P.O: Sainik School, Bhubaneswar - 751 005, Odisha, India}
\address[label3]{Variable Energy Cyclotron Centre, 1/AF Bidhan Nagar, Kolkata-700 064, West Bengal, India}
\address[label4]{Department of Applied Physics, CU, 92, APC Road, Kolkata-700 009, West Bengal, India }

\begin{abstract}
At NISER-IoP detector laboratory an initiative is taken to build and test Gas Electron Multiplier (GEM) detectors for ALICE experiment. The optimisation of the gas flow rate and the long-term stability test of the GEM detector are performed. The method and test results are presented.   
\end{abstract}
\begin{keyword}
LHC \sep ALICE \sep Gas Electron Multiplier \sep Long-term test

\end{keyword}
\end{frontmatter}

\section{Introduction}\label{intro}
Gas Electron Multiplier (GEM) is one of the most important micro-pattern gaseous detectors used in the recent and being considered for future High-Energy Physics (HEP) experiments \cite{FS97, ALICE}. For example A Large Ion Collider Experiment (ALICE) at the Large Hadron Collider (LHC) Facility is upgrading its multi-wire proportional chamber based Time Projection Chamber (TPC) with GEM units. In line with the worldwide efforts, we have also taken an initiative in NISER-IoP experimental high-energy physics detector laboratory to build and test of GEM detector prototypes. The GEM foils and other components are obtained from CERN \cite{GDD}. The detector is tested with Co$^{60}$, Cs$^{137}$ and Sr$^{90}$ radioactive sources with Ar/CO$_2$ gas in 70/30 volume ratios. The temperature and atmospheric pressure are measured and recorded continuously using a data logger developed in-house \cite{Sahu}. Effect of the temperature and pressure on anode current is measured. Variation of the count rate is also measured with varying flow rate. The long-term stability test is performed measuring the anode current with time. The details test results are presented in this article.

\section{Description of the GEM module}\label{construct}
A triple GEM detector prototype, consisting of 10~cm~$\times$~10~cm standard stretched single mask foils, obtained from CERN is assembled with the drift gap, 2-transfer gaps and induction gap of 3,2,2,2~mm respectively. A voltage divider network is built by resistors and a single high voltage (HV) channel is used to power the GEM detector. The detector has XY printed board (256 X-tracks, 256 Y-tracks) in the base plate and worked as readout plane. Each of 256 X-tracks and 256 Y-tracks are divided into two 128 pin connectors. In each 128 pin connector a sum-up board (provided by CERN) is used and a single signal is taken by a Lemo cable. The signal is fed to a charge sensitive pre-amplifier and subsequently to a spectroscopic amplifier. After discrimination the signals are counted by a scaler.

On the other hand during the anode current measurement the Lemo output of the sum-up board is directly connected by a very small Lemo cable to a 6485 Keithley Pico-ammeter.

\section{Experimental results}
In this section the experimental results are presented that mainly include optimisation of the gas-flow rate and long-term stability test.
\subsection{Optimisation of gas-flow rate}
The variation of the count rate with gas flow rate is measured. The gas flow rate is measured using the  water displacement method. A measuring cylinder is filled with water and its top is covered with a plastic sheet. It is then inverted in a plastic box half-filled with water such that the plastic foil was lying completely flat on the base of the box, but still covering the top of the measuring cylinder. The plastic cover is gently pulled out and the open end of the gas-outlet tube from the detector is now gently put into the measuring cylinder by slightly tilting it. A particular volume (70~ml) of the gas in the cylinder is measured and the time interval (t in minutes) is measured using a stop watch. The gas-flow rate is calculated. 


\begin{figure}[htb!]
\begin{center}
\includegraphics[scale=0.4]{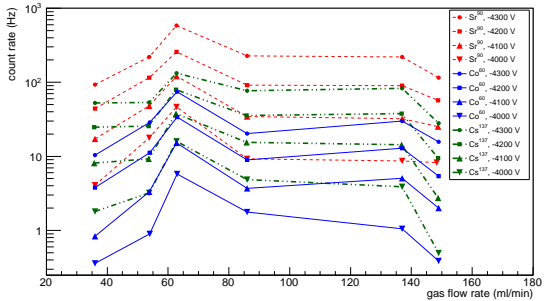}
\caption{\label{flow}Count rates vs. gas flow rate for $Co^{60}$, $Cs^{137}$ and $Sr^{90}$.}\label{flow}
\end{center}
\end{figure}
\begin{figure}[htb!]
\begin{center}
\includegraphics[scale=0.4]{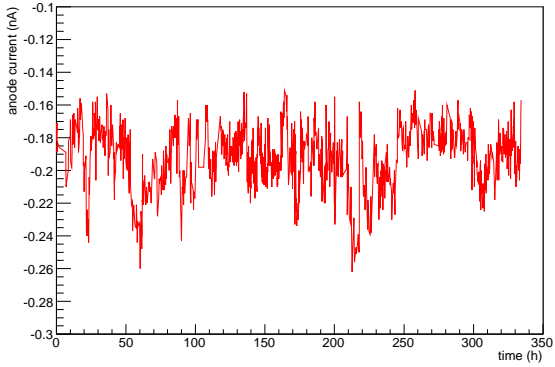}
\caption{Variation of anode current as a function of time.}\label{anodetime}
\end{center}
\end{figure}
\begin{figure}[htb!]
\begin{center}
\includegraphics[scale=0.4]{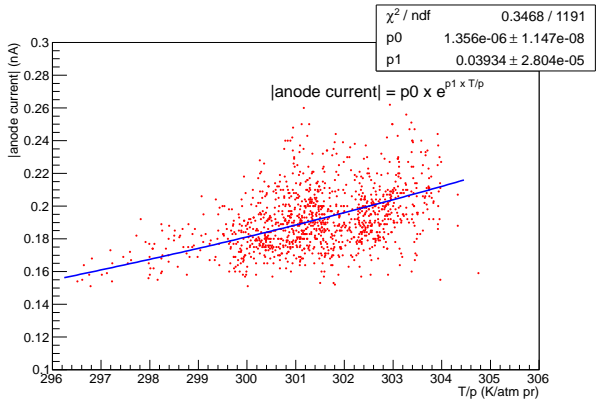}
\caption{$|$ anode current $|$ as a function of T/p.}\label{anodeTbyp}
\end{center}
\end{figure}
\begin{figure}[htb!]
\begin{center}
\includegraphics[scale=0.4]{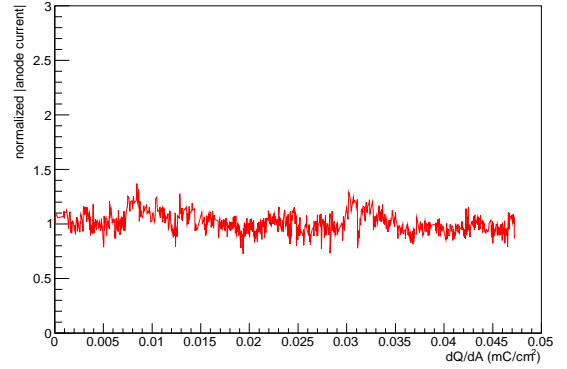}
\caption{Variation of normalised $|$anode current$|$ as a function dQ/dA.}\label{normvscharge}
\end{center}
\end{figure}

The count rates are measured with different gas flow rates and different voltages. The plots obtained for count rate as a function of the gas flow rate for different sources, at different applied HV as shown in Figure$~\ref{flow}$. From Figure$~\ref{flow}$ the count rates are found to be maximum at around 65 ml/min gas flow rate for all sources and for all applied voltages.

\subsection{Long term stability test}
The long-term stability of the triple GEM detector is studied using $Sr^{90}$ source and measuring the anode current with and without source continuously \cite{SB12}. At some intervals of 10 minutes, the anode current with and without source are measured. Simultaneously the temperature (t in $^oC$), pressure (p in mbar) and relative humidity (RH in $\%$) are recorded from a data logger, built in-house, with a time stamp \cite{Sahu}.
The output anode current for the source is given by,
\begin{equation}
i_{source} = i_{with~source} - i_{without~source}
\end{equation}
where $i_{source}$ is anode current due to source, $i_{with~source}$ is the measured anode current when the detector is irradiated by the source and $i_{without~source}$ is the anode current without any source. The variation of anode current due to source is plotted as a function of total period of operation in Figure~$\ref{anodetime}$. 

The $|$ anode current $|$ is plotted as a function of T/p and fitted with a function
\begin{equation}
|anode current|(T/p) = Ae^{B(T/p)}
\end{equation}
and is shown in Figure~$\ref{anodeTbyp}$, where T (= t+273) is the absolute temperature in Kelvin and p is in atmospheric pressure.

It is well known that gain of a gas detector depends on T/p. The relation between the gain (G) of a GEM detector on absolute temperature and pressure is given by \cite{Altunbas}
\begin{equation}
G(T/p) = Ae^{B(T/p)}.
\end{equation}


The value of the parameters A and B obtained are 1.356~$\times$~10$^{-6}$ and 0.03934 atm pr/K. Using the fit parameters, the $i_{source}$ was normalised by using the following relation:
\begin{equation}
i_{normalized} = \frac{i_{source}}{Ae^{B(T/p)}}
\end{equation}

To check the stability of the detector with continuous radiation, the normalised $|$ anode current $|$ was plotted against the total charge accumulated per unit irradiated area (that is directly proportional to time) of the detector. To calculate the total charge accumulated, the average current of two time say $t_1$ and $t_2$ is taken and multiplied by the time interval $(t_2-t_1)$. The total charge accumulated will be the sum of accumulated charge over all the intervals during every two adjacent readings. To get total charge accumulated per unit area, the total charge accumulated was divided by the area of the irradiated area. The normalised $|$ anode current $|$ as a function of dQ/dA is shown in Figure~$\ref{normvscharge}$. The distribution of the normalised $|$anode current$|$ when fitted with a Gaussian function the mean has been found to be around 1 with a sigma of 0.079.

\section{Conclusions and outlooks}
Triple GEM detector prototype building is started. The detector is tested with a gas mixture of Ar/CO$_2$ of 70/30 volume ratio. The count rate of the detector depends on the gas flow rate and it is optimised. This is because due to increase of gas flow rate the gain of the detector increases if there is leak in the detector somewhere. Because increasing gas flow rate the electronegative O$_2$ content decreases \cite{SB11}. On the other hand increasing gas flow rate further the pressure increases which decreases the gain. When this two effects work simultaneously then a optimum flow rate is observed which gives the maximum gain.

The long-term stability test of this detector is performed using $Sr^{90}$ beta radioactive source. No ageing is observed even after operation of the GEM detector for about 350 hours or after an accumulation of charge per unit area of 0.05 mC/cm$^2$.

\section{Acknowledgements}
We would like to acknowledge Mr. Rudranarayan Mohanty and Mr. Lalita M. Jena for their help in the laboratory work. We also acknowledge Mr. Md. Rihan Haque for writing a code for the data analysis. We would like to thank Dr. Chilo Garabatos for his valuable discussions and suggestions during course of the work. S. Biswas acknowledges the support of DST-SERB Ramanujan Fellowship (D.O. No. SR/S2/RJN-02/2012). XII$^{th}$ Plan DAE project titled Experimental High Energy Physics Programme at NISER-ALICE is also acknowledged.

\end{document}